\documentstyle[11pt,newpasp,twoside,epsf]{article}
\def\edcomment#1{\iffalse\marginpar{\raggedright\sl#1\/}\else\relax\fi}
\reversemarginpar

\begin{document}
\title{Three New Binary Pulsars Discovered With Parkes}
\author{Jason Hessels$^1$, Scott Ransom$^{1,2}$, Mallory Roberts$^{1,2}$, Victoria Kaspi$^1$, 
Margaret Livingstone$^1$, Cindy Tam$^1$, and Fronefield Crawford$^3$}

\affil{$^1$Department of Physics, McGill University, 3600 rue Universit\'e, Montr\'eal, Qu\'ebec, Canada, H3A-2T8.} 
\affil{$^2$Center for Space Research, MIT, 70 Vassar St., Bldg. 37, Cambridge, MA 02139.} 
\affil{$^3$Department of Physics, Haverford College, Haverford, PA 19041.} 

\begin{abstract}
We present three new binary pulsars discovered during a search for pulsations
in 56 unidentified mid-latitude EGRET $\gamma$-ray error boxes with the Parkes
 multibeam receiver.
Timing observations of these sources is on-going with both the Parkes and the Green Bank
 telescopes.
We discuss the place of these new systems in the population of binary pulsars and suggest
 that they are all somewhat atypical systems.
\end{abstract}


\section{Introduction}

We have recently completed a search of 56 unidentified mid-latitude 
EGRET $\gamma$-ray sources for pulsations using the Parkes telescope and the multibeam receiver.  
This survey covered $\sim$140 square degrees of sky at Galactic latitudes $> 5^{\circ}$
and revealed six new pulsars, three of which are in binary systems.  
Here we present initial timing results for the three binaries.


\section{The Pulsars}

The properties of the three new binaries are summarized in Table~1
 and their pulse profiles are shown in Figure~2.


\begin{table}
\caption{Newly Discovered Binaries}
\begin{tabular}{l c c c c c c}
\hline \hline
Pulsar       &  $P_{psr}$ & $P_{orbit}$ & $a_1 \sin (i)/c$ & Min $M_2^a$ & DM             & Distance$^b$ \\
             & (ms)       & (days)      & (lt-s)           & ($M_{\sun}$) & (pc cm$^{-3}$) &  (kpc)      \\
\hline 
J1614$-$2230 &  3.151     & 8.7         & 11.3             & 0.40        & 34.5           &   1.3        \\
J1614$-$2318 &  33.50     & 3.2         & 1.33             & 0.080       & 52.4           &   1.8        \\
J1744$-$3922 &  172.4     & 0.19        & 0.212            & 0.083       & 148            &   3.1        \\
\hline \hline
\end{tabular}
$^a$ Assuming a pulsar mass ($M_1$) of 1.4\,M$_{\sun}$. \\
$^b$ DM distance from the NE2001 electron density model of the Galaxy (Cordes \& Lazio 2002, astro-ph/0207156). \\
\end{table}

\begin{figure}[!t]
\plotone{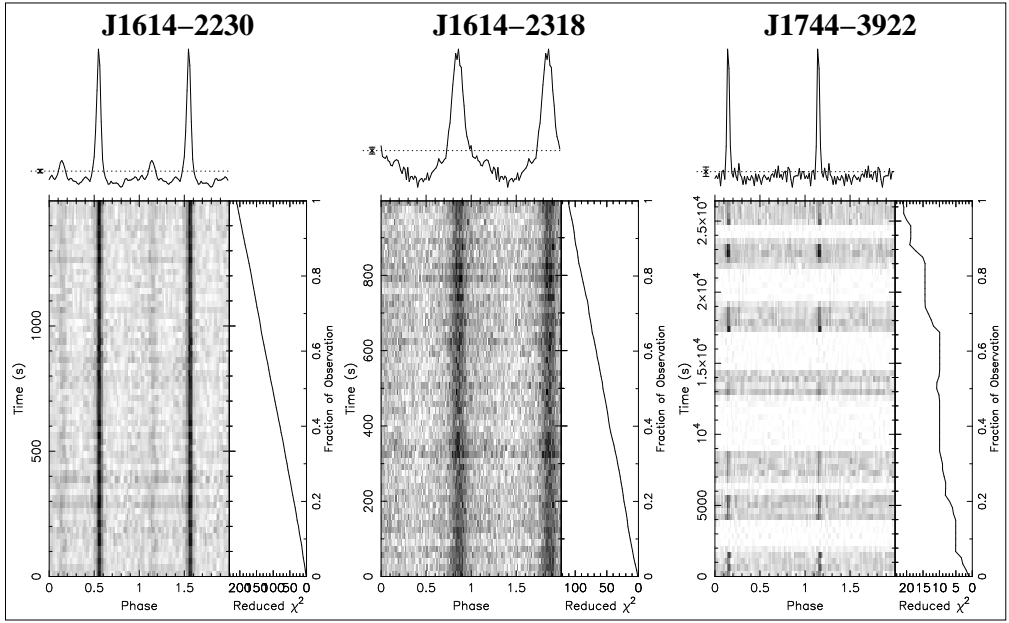}
\caption{Best detections of the pulsars.  The grey scale shows the consistency
 of the pulsations as a function of time during the observation and effects
 due to interference.  
{\it Left:} PSR~J1614$-$2230 with the GBT using the BCPM at 1.4\,GHz.  
{\it Middle:} PSR~J1614$-$2318 with the GBT using the BCPM at 350\,MHz.
{\it Right:} PSR~J1744$-$3922 with Parkes using the multibeam filterbank 
at 1.4\,GHz.  The profile is made up of numerous
 closely spaced observations, covering $\sim$1.5 orbits of the pulsar, 
with padding in between.}
\end{figure}

\subsection{PSR~J1614$-$2230}

PSR~J1614$-$2230 is a recycled millisecond pulsar with a very low inferred magnetic field ($B = 1.8 \times 10^{8}$\,G). 
It has the highest minimum companion mass of the $\sim$50 binary pulsars with 
spin periods lower than 8\,ms, suggesting a possible non-standard evolution.
Compared with the orbital period versus companion mass relationship of 
 Rappaport et al. 1995 (MNRAS, 273, 731) the orbital period 
of PSR~J1614$-$2230 is short by a factor $>$10 for its companion mass.
The companion may be a CNO white dwarf or a low-mass degenerate dwarf. 
The spin-down rate of PSR~J1614$-$2230 gives an 
$\dot{E} \sim 1.3 \times 10^{34}$ erg s$^{-1}$, which is just barely consistent
 with the flux measured from the $\gamma$-ray source in whose error box this pulsar was found.  If the pulsar is the counterpart it would demand a very
 high $\gamma$-ray efficiency, or a significantly closer distance than what is inferred from the DM.

\newpage
\subsection{PSR~J1614$-$2318}

PSR~J1614$-$2318 is likely partially recycled ($B \sim 3 \times 10^{9}$\,G).  
It has the lowest minimum companion mass of any binary pulsar with spin period 
between 10 and 100\,ms.  The pulse profile is relatively wide at 1.4\,GHz 
($\sim$30\% of the pulse period at FWHM).  At $\sim$350\,MHz
 the higher signal-to-noise ratio of the profile shows that there are wings on either side
 of the main pulse (see Figure~1).  
It is possible that we are seeing emission from the core and cone of the beam.

\subsection{PSR~J1744$-$3922}

PSR~J1744$-$3922 was also independently discovered by the Parkes Multibeam Galactic Plane Survey 
 (A. Lyne, private communication).
It is possibly partially recycled ($B \le 7 \times 10^{10}$\,G) and is one 
of only $\sim$16 binary pulsars with spin periods greater 
than 50\,ms (of these it has the shortest orbital period and the third smallest minimum companion mass).
It is difficult to detect at times, as it seems to shut off and
 turn on on timescales as short as a few tens of seconds (see Figure~2).
Since its DM is relatively high, it is unlikely that scintillation is responsible for
 the many non-detections of the pulsar.  PSR~J1744$-$3922 has also been very difficult
 to detect at wavelengths other than $\sim$1.4\,GHz.  Observations with the GBT have improved
 the detection rate of this strange pulsar.

\begin{figure}[!t]
\plotone{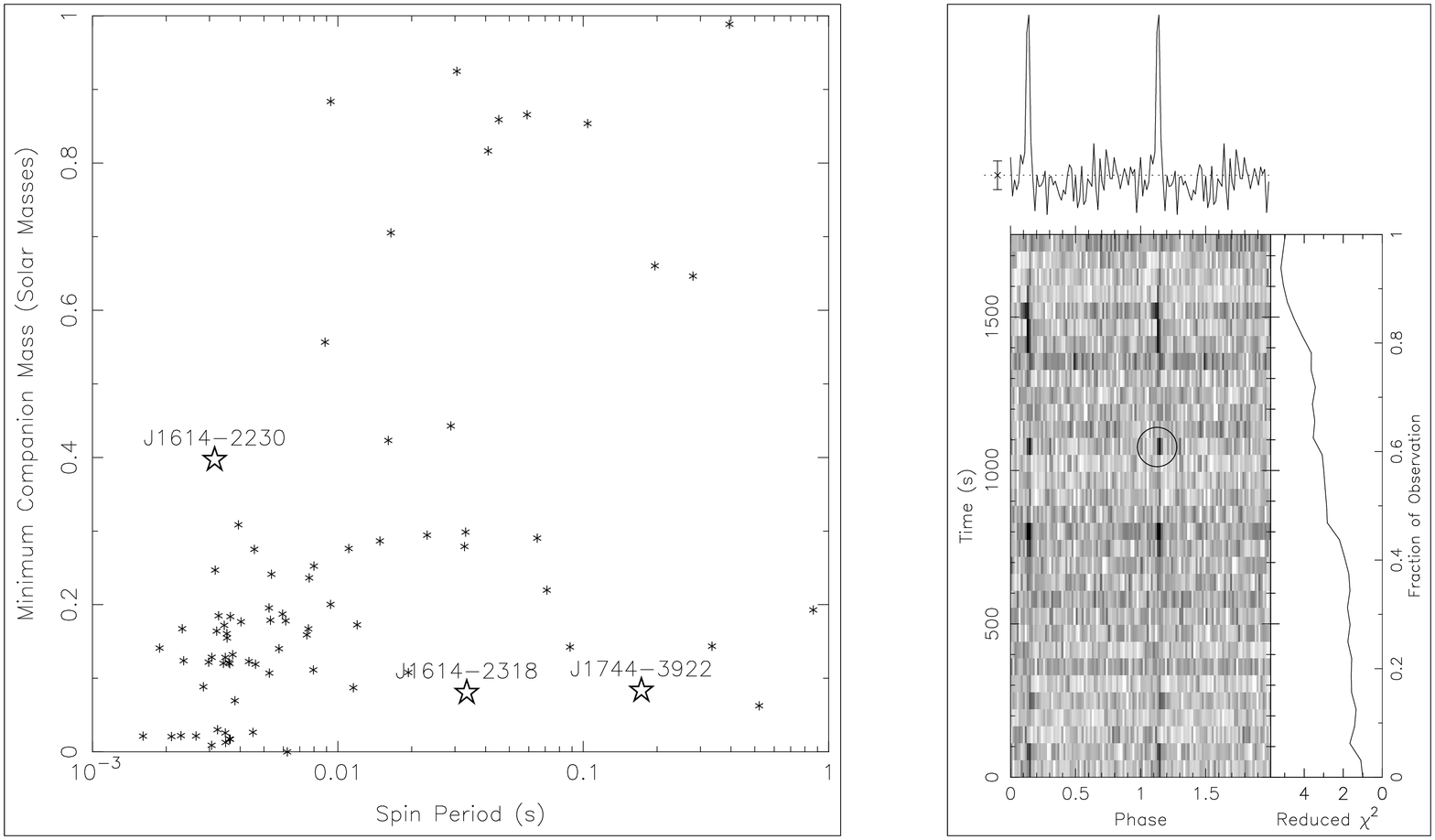}
\caption{{\it Left:} A plot of minimum companion mass versus spin period for the
known binary pulsars (from the ATNF catalog, available at http://www.atnf.csiro.au/research/pulsar/psrcat/).  
The three binaries presented here are marked with a star
 and their name.  {\it Right:} A Parkes observation of PSR~J1744$-$3922 at $\sim$1.4\,GHz
 showing the intermittent nature of the pulsations from this source.  A particularly
short episode ($<$70\,s) of relatively bright pulsations is circled.}
\end{figure}

\section{Future Work}

We have applied for exploratory observations of all three pulsars presented here with
 the 8-m Gemini-South optical telescope in order to determine the magnitudes and colors
 of their binary companions and to see if they are consistent with being white dwarfs.  
These observations may lead to future spectroscopic 
 observations of the companions, which could allow us to determine the neutron star masses and distances to the systems.
With a clearer picture of the nature of the companion stars we would also be in a better
 position to comment on how these systems fit in with theories of binary evolution.

\end{document}